\documentclass{aastex}
\usepackage{emulateapj5}

\begin{document}

\title{UV excess galaxies: Wolf-Rayet galaxies}
\author{Warren R.~Brown, Scott J.~Kenyon, Margaret J.~Geller, and Daniel
G.~Fabricant}
\affil{Harvard-Smithsonian Center for Astrophysics, 60 Garden St.,
Cambridge, MA 02139}

\email{wbrown@cfa.harvard.edu,
skenyon@cfa.harvard.edu,
mgeller@cfa.harvard.edu, 
and dfabricant@cfa.harvard.edu
}

\shorttitle{Wolf-Rayet Galaxies}
\shortauthors{Brown et al.}

\begin{abstract}

	We discuss $V$ and $R$ band photometry for 67\% of the
\citet{S2000} SA57 ultraviolet-selected galaxy sample.  In a sample
of 176 UV-selected galaxies, \citet{S2000} find that 24\% have (UV-$B$)
colors too blue for consistency with starburst spectral synthesis models.
We propose that these extreme blue, UV excess galaxies are Wolf-Rayet (WR)
galaxies, starburst galaxies with strong UV emission from WR stars. We
measure a median $(\vr)=0.38\pm0.06$ for the UV-selected sample,
bluer than a sample optically selected at $R$ but consistent with
starburst and WR galaxy colors.  We demonstrate that redshifted WR
emission lines can double or triple the flux through the UV bandpass at
high redshifts.  Thus the (UV$-B)$ color of a WR galaxy can be up to 1.3
mag bluer at high redshift, and the expected selection function is skewed
to larger redshifts.  The redshift distribution of the extreme blue, UV
excess galaxies matches the selection function we predict from the
properties of WR galaxies.

\end{abstract}

\keywords{galaxies: starburst, galaxies: photometry, stars: Wolf-Rayet,
ultraviolet: galaxies}

\section{INTRODUCTION}

	\citet[hereafter S2000]{S2000} published an ultraviolet (UV)
flux-limited catalog of galaxies, imaged with the FOCA balloon-borne UV
camera of \citet{mil92}.  The catalog is an extension of work by
\citet{tre98}, and contains 273 galaxies with redshifts $z\lesssim0.4$.
Vacuum UV imaging enables the direct detection of radiation from hot,
massive stars and, by association, actively star forming galaxies. The
luminosity function (LF) of the UV-selected galaxies is somewhat
surprising.  S2000 find a faint end slope in the \citet{sch76}
parametrization of the LF, $\alpha_{uv}=-1.55\pm0.11$, steeper than
typical optical LF faint end slopes \citep[etc.]{fol99, mar98, gel97}.  
As a result, S2000 find no strong decrease in UV luminosity density
between $z\simeq0.4$ and $z=0$ and thus little decline in the density of
star formation over this range of cosmic epochs.

	In contrast, \citet{mad96} conclude that there is a factor of
$\simeq10$ decline in the comoving density of star formation between
redshifts $z=1$ and $z=0$. This conclusion is based on a combination of
rest frame near-UV luminosities from the $I$-selected Canada France
Redshift Survey \citep{lil95} and local H$\alpha$ measures from Gallego et
al.'s (1995) objective prism survey.  

	Curiously, a substantial fraction of the S2000 UV-selected
galaxies have (UV$-B)$ colors bluer than the bluest color that can be
produced by starburst spectral synthesis models \citep{fio97,lei99}. S2000
could therefore not explain the color of these galaxies.

	Here we propose that the extremely blue galaxies are Wolf-Rayet
(WR) galaxies.  WR galaxies show direct signatures of WR stars in their
spectra, usually a broad He {\small II} $\lambda4686$ feature originating
in the stellar winds of WR stars.  This subtle feature is not seen in the
fiber-spectra of S2000 (M. Sullivan 2000, private communication), but
requires good S/N and is spatially dependent on the inclusion of starburst
regions \citep{sch99}.  The starburst regions dominate the broadband UV
flux and, though short-lived ($<10^{7}$ yrs), WR stars can dominate the UV
light of young starbursts.

	In this paper we describe optical photometry for 67\% of the S2000
Selected Area 57 (SA57) UV-selected galaxy sample.  Our goal is to compare
the optical and UV colors, and to suggest that WR emission lines are the
source of the extremely blue colors found by S2000.  In section 2 we
discuss our data, in section 3 we discuss the (\vr) colors, and in section
4 we discuss the impact of Wolf-Rayet UV emission lines on the
interpretation of the UV sample.  We conclude in section 5.

\section{DATA}

	We obtained broadband $V$ and $R$ images in 1998 December and 1999
February with the MOSAIC camera on the Kitt Peak National Observatory 0.9m
telescope.  MOSAIC is an 8 CCD array with pixel scale 0.424 arcsec
pixel$^{-1}$ and field of view $59^\prime $ x $59^\prime $ on the 0.9m. We
obtained three offset 200 sec $R$ and 333 sec $V$ images at each position
along a strip in RA.  Our data serendipitously overlap with 148 of the 222
S2000 UV-selected galaxies in the SA57 region.

	We processed the images in the standard way with the IRAF MSCRED
package.  Total magnitudes are obtained using SExtractor \citep{ber96}.
Seven percent of the magnitudes are obtained with a modified version of
IRAF-based GALPHOT \citep{fre93}.  GALPHOT allows photometry of galaxies
poorly fit by SExtractor, including low surface brightness galaxies or
galaxies with close companions.  A comparison between the two software
packages yields a 1-$\sigma$ scatter in the total galaxy magnitudes of
$\pm0.05$ mag.  We use Landolt fields \citep{lan92} and the M67 cluster
\citep{mon93} for flux calibration.  Depending on the nightly seeing
conditions, the error in the standard star calibration is $\pm0.02-0.03$
mag.  A direct measure of our intrinsic error is the RMS scatter of
measurements of identical galaxies taken on different nights: $\pm0.04$
mag.

	The photometry includes galactic extinction corrections from
\citet{sch98}, who combine $COBE$/DIRBE and $IRAS$/ISSA data to map dust
emission in the Milky Way.  Typical extinction values are $A_R=0.025$ and
$A_V=0.03$ mag.  The photometry also includes ``k corrections'' for the
changing rest frame bandpass \citep{fre94}.  Because we know the (\vr)
color and the redshift of each galaxy, we can determine accurate k
corrections. Typical values for the $V$ and $R$ bandpasses are $-0.1 < k_R
< 0.1$ and $-0.15 < k_V < 0.15$ mag.

\subsection{Existing UV and $B$ measurements}

	The S2000 UV measurements were made from a balloon-borne, 40 cm
Cassegrain telescope stabilized to 2 arcsec RMS.  The filter response
approximates a Gaussian centered at 2015 \AA, with a FWHM of 188 \AA.  
The limiting magnitude in the UV catalog is m$_{UV}=18.5$.  S2000
state that the UV zero-point is accurate to $\le 0.2$ mag; the uncertainty
in the relative photometry may reach $\simeq 0.5$ mag at the limiting
magnitude.

	With the UV images in hand, S2000 searched for optical
counterparts in digitized POSS plates.  S2000 obtained accurate positions
and $B$ photometry, with uncertainty $\pm0.20$ mag, from the POSS plates.
We follow S2000 and only use galaxies with a single optical counter-part
when discussing UV colors.  S2000 used the resulting (UV$-B)$ colors to
calculate k corrections.  Finally, S2000 apply a reddening correction
based on H$\alpha$ and H$\beta$ line strengths and the empirical
\citet{cal97} extinction law for starburst galaxies.  They applied a mean
reddening correction $A_V=0.97$ to galaxies lacking H$\alpha$ and/or
H$\beta$ emission.

	The (UV$-B)_{\rm o}$ colors we quote from S2000 include k
corrections and reddening corrections.  Our (\vr) colors include k
corrections but no reddening corrections, because S2000 do not provide
enough information to reproduce their reddening corrections.  Even if this
reddening data were provided, 72\% of the galaxies with (\vr) colors lack
both H$\alpha$ and H$\beta$ emission.  For these galaxies, the mean
reddening correction would add a uniform $\simeq-0.2$ mag offset to the
(\vr) colors.

\section{OPTICAL COLORS}

	Figure 1 shows that the (\vr) colors of the UV-selected galaxies
are significantly bluer than those from an $R$-selected galaxy sample, the
Century Survey \citep{gel97}.  The Century Survey is a complete redshift
survey of 102 square degrees to a limiting magnitude $R=16.13$, containing
1762 galaxies.  The Century Survey galaxies have median
$(\vr)_{CS}=0.58\pm0.04$; the UV-selected galaxies have median
$(\vr)=0.38\pm0.06$.

	The distribution in Figure 1 appears double-peaked, but
application of the non-parametric estimator of \citet{pis93} shows the
second peak is not significant.  The \citet{pis93} kernel probability
estimator allows estimation of the probability density distribution
underlying a data sample.  In applying this we assume an average
1-$\sigma$ (\vr) error $\pm0.066$ mag.

	The S2000 (UV$-B)_{\rm o}$ colors provide a measure of the star
formation activity, but there is no correlation between the (\vr) and
(UV$-B)_{\rm o}$ colors.

	Unlike the extreme blue (UV$-B)_{\rm o}$ colors found by S2000,
however, the optical (\vr) colors are consistent with starburst galaxy
colors.  For example, the template starburst-galaxy spectral energy
distributions SB1 and

        \includegraphics*[scale=0.44]{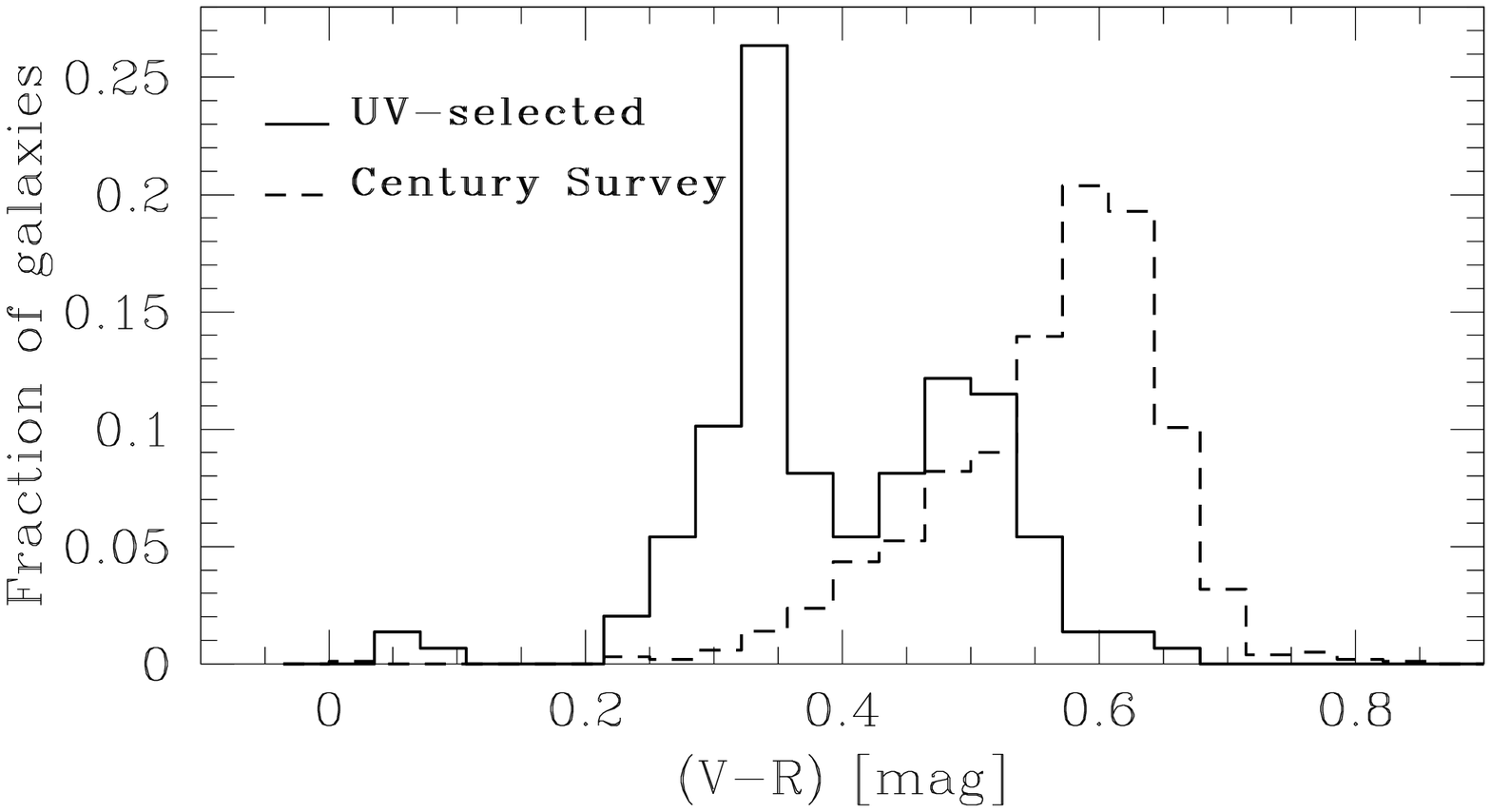}
 
\figcaption[fig1b.ps]{Histogram of (\vr) colors for UV-selected galaxies  
(solid line) and $R$-selected Century Survey (dashed line).}

	~

\noindent SB2 of \citet{kin96} have synthetic $(V-R)=0.26$.  All but
two of the S2000 UV-selected galaxies are redder than $(V-R)=0.23$, and we
note that these blue outliers appear atypically faint and point-like in
our images.  Thus $\sim95\%$ of the UV galaxy (\vr) colors are consistent
with starburst galaxy colors. If the extreme blue galaxies were AGN, their
average color should be $(\vr)=0.60$ \citep{kot93}, which is clearly not
the case.

\section{DISCUSSION}

	An unresolved issue from the S2000 paper is the existence of
galaxies with (UV$-B)_{\rm o}$ colors $\lesssim-2$, colors too blue to be
explained by standard spectral synthesis models.  In our subsample of the
S2000 UV-selected galaxies, 48 of 148 galaxies have (UV$-B)_{\rm o}<-2$,
yet the optical (\vr) colors show no significant departure from the rest
of the sample.  A KS test gives an 80\% probability of the extreme
(UV$-B)_{\rm o}<-2$ and the other UV-selected galaxies being drawn from
the same distribution of (\vr) colors.

	Next we examine the redshift distribution of the extremely blue
galaxies.  The lower panel of Figure 2 shows the redshift distribution of
the extreme blue (UV$-B)_{\rm o}<-2$ galaxies, and the upper panel the
remaining (UV$-B)_{\rm o}>-2$ galaxies.  Figure 2 shows that the extreme
blue galaxies are preferentially found at high redshift. A KS-test yields
a $10^{-11}$ probability of the two populations being drawn from the same
distribution.

	The peaks in the redshift histograms (Figure 2) are well matched
to the $z\sim0.3$ limit of the Medium Deep Survey \citep{wil96}, which
includes the SA57 region.  The Medium-Deep Survey is $\sim70\%$ complete
to $B_J\sim20.5$ mag, and gives a sense of the existing large scale
structure in the direction of UV survey.  Thus the extremely blue galaxies
appear to trace the same large-scale structure as ``normal'' galaxies.

\subsection{Wolf-Rayet galaxies.}

	The very blue (UV$-B)_{\rm o}<-2$ galaxies are an extreme
population of galaxies found mostly at larger redshift.  An explanation
probably requires emission lines redshifting into the UV bandpass.  
Because S2000 and our $(\vr)$ color distribution rule out AGN, it seems
plausible the emission lines come from a stellar emission line source,
i.e.~Wolf-Rayet (WR) stars, planetary nebulae, or symbiotic stars.  
Because there are known galaxies with WR features, we suggest that the
extreme blue (UV$-B)_{\rm o}<-2$

        \includegraphics[scale=0.43]{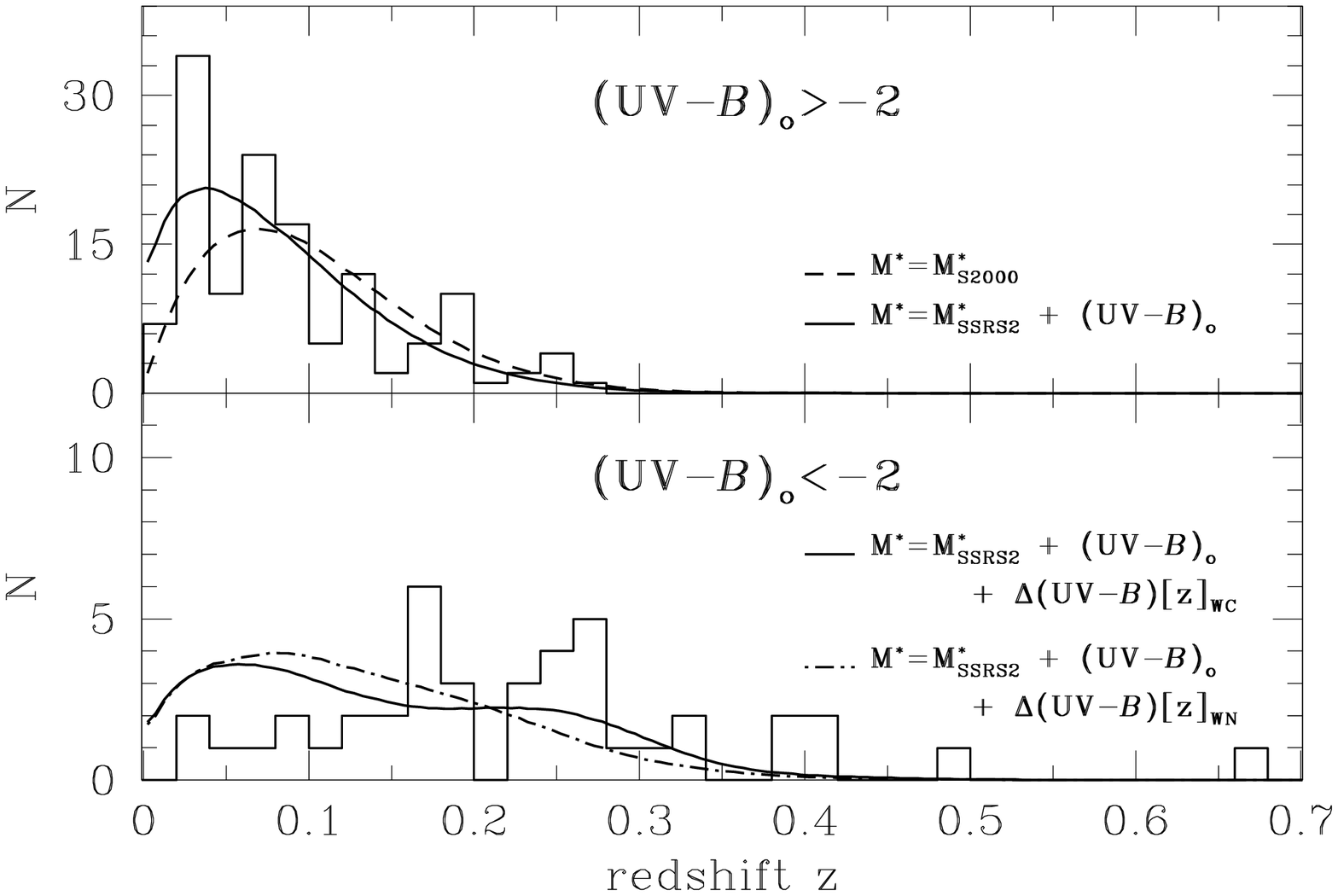}
 
\figcaption[fig2b.ps]{Redshift histograms for the (UV$-B)_{\rm o}>-2$  
galaxies (upper panel) and the extreme blue (UV$-B)_{\rm o}<-2$ galaxies
(lower panel). Note the excess of extreme blue galaxies at large
redshift.  The selection function curves are calculated with the S2000 LF
parameters (dashed-line) and the SSRS2 Sm-Irr LF parameters
(dash-dot, solid lines) as described in section 4.1.}

	~

\noindent galaxies are WR galaxies.  The strongest
UV features come from ionized C, O, He, Si, and Fe in the stellar winds of
WR stars (e.g.~Hillier \& Miller 1999).

	Our optical spectra of 10 extreme blue galaxies all show narrow
emission lines indicative of active star formation.  WR galaxies are
characterized by the direct signatures of WR stars in their spectra, but
our S/N is too poor to detect a possible $\lambda4686$ WR feature
convincingly.

	To calculate a typical (\vr) color for WR galaxies, we searched
the NASA/IPAC Extragalactic Database for photometery of known WR galaxies.  
The average optical color of 13 WR galaxies is
$\overline{(\vr)}=0.27\pm0.34$ mag.  This average color is in agreement
with the blue color we find for the (UV$-B)_{\rm o}<-2$ sample,
$\overline{(\vr)}=0.35\pm0.11$ mag.

	Figure 3 illustrates how WR emission lines are redshifted into the
S2000 UV bandpass.  We plot a sample WR ultraviolet spectrum, WC star HD
165763 \citep{hil99}, at redshift $z=0$ and at redshift $z=0.285$ with the
S2000 UV and $B$ bandpasses overlaid. Integration of the UV bandpass over
the redshifted spectrum shows that the WR emission lines can boost the UV
flux by a factor of $\lesssim3$ compared to the flux at $z=0$.  We also
integrate the $B$ bandpass over the \citet{hil99} WC stellar spectrum, and
find the (UV$-B)$ color is shifted up to 1.3
mag bluewards over the redshift range z=0 to z=0.5 (see

	\includegraphics*[scale=0.43]{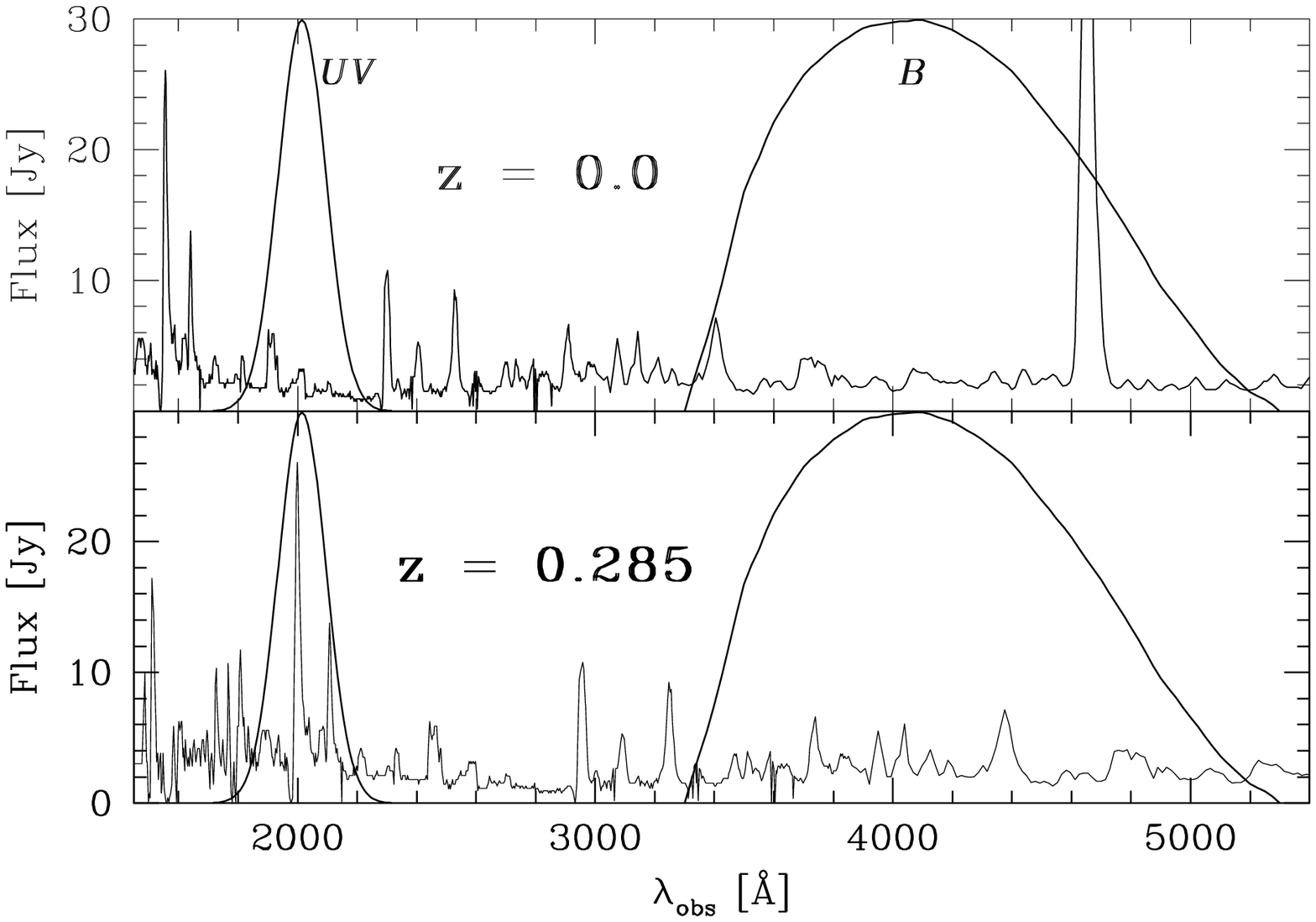}
 
\figcaption[fig3b.ps]{WR spectrum of star HD 165763 at $z=0$ and at
$z=0.285$, with rest-frame UV and $B$ bandpasses.}

\noindent Figure 4), blue enough to explain $2/3$ of the extreme blue
(UV$-B)_{\rm o}$ galaxies in the SA57 UV-selected sample.

	The WC stellar spectrum represents a maximal effect; known WR
galaxies have typical WC/WN ratios of 0.2 - 0.4 \citep{sch99}.  We perform
the (UV$-B)$ calculation on the WN stellar spectrum, WR152 \citep{ham98},
and find the (UV$-B)$ color shifts up to 1.1 mag bluewards over the same
redshift range (see Figure 4).

	We acknowledge that strong UV emission lines are not common
features of known low redshift WR galaxies (Leitherer 2000, private
communication).  We calculate the shifts in (UV$-B)$ for a dozen of the
\citet{mcq95} ``UV to optical spectral distributions of northern
star-forming galaxies,'' and find an average $\Delta({\rm UV}-B)\sim-0.3$
mag (mostly due to the continua, not the emission lines). Yet good
examples of the very blue objects may not be found locally.  The UV
selection should pick out extreme objects; known WR galaxies are optically
selected. We note the very blue UV galaxies have median z=0.2. By
comparison, the \citet{sch99b} catalog of known WR galaxies drops off
sharply at z=0.02.

	We can also take advantage of $HST$ FOS observations of the Small
 Magellanic Cloud H {\small II} region N88A \citep{kur99} as a UV
 spectral templates for WR galaxies.  When the S2000 bandpass is centered
 on the strong C{\small III} and Si{\small III} lines in this object, the
 UV flux doubles compared to the surrounding continuum level.

	S2000 compute their luminosity function (LF) with the full UV
sample; here we construct a model LF for the WR galaxies alone.  The
impact of the emission lines on the observed UV flux changes with
redshift.  We explore the selection function for finding WR galaxies in a
magnitude limited redshift survey by computing the change in absolute
magnitude resulting from the UV emission lines.

	WR galaxies are blue and show active star formation.  We thus use
the Second Southern Sky Redshift Survey \citep[SSRS2]{mar98} LF for the
bluest galaxies, the Magellanic spiral and irregulars, as the basis of our
model. The SSRS2 is a $B$-selected survey complete to $m_b\le15.5$
that contains 5404 galaxies, all with morphological classifications.  The
SSRS2 Sm-Irr LF has a steep faint-end slope, $\alpha=-1.81$, similar to
the S2000 LF, $\alpha=-1.55$.

	We compute the selection function for the WR galaxies according to
equation (1),
 \begin{equation}
N(<M_{lim}) \propto \int_z D_C(z)^2 dD_C(z)
\int_{M=-\infty}^{M_{lim}(z)} \phi(M,z) dM
 \end{equation}
	where $D_C(z)$ is the co-moving distance.  We assume a flat
cosmology and $H_o$=100 km s$^{-1}$ Mpc$^{-1}$, $q_o$=0.5.  $\phi(M,z)$ is
the luminosity function in the \citet{sch76}

	\includegraphics*[scale=0.43]{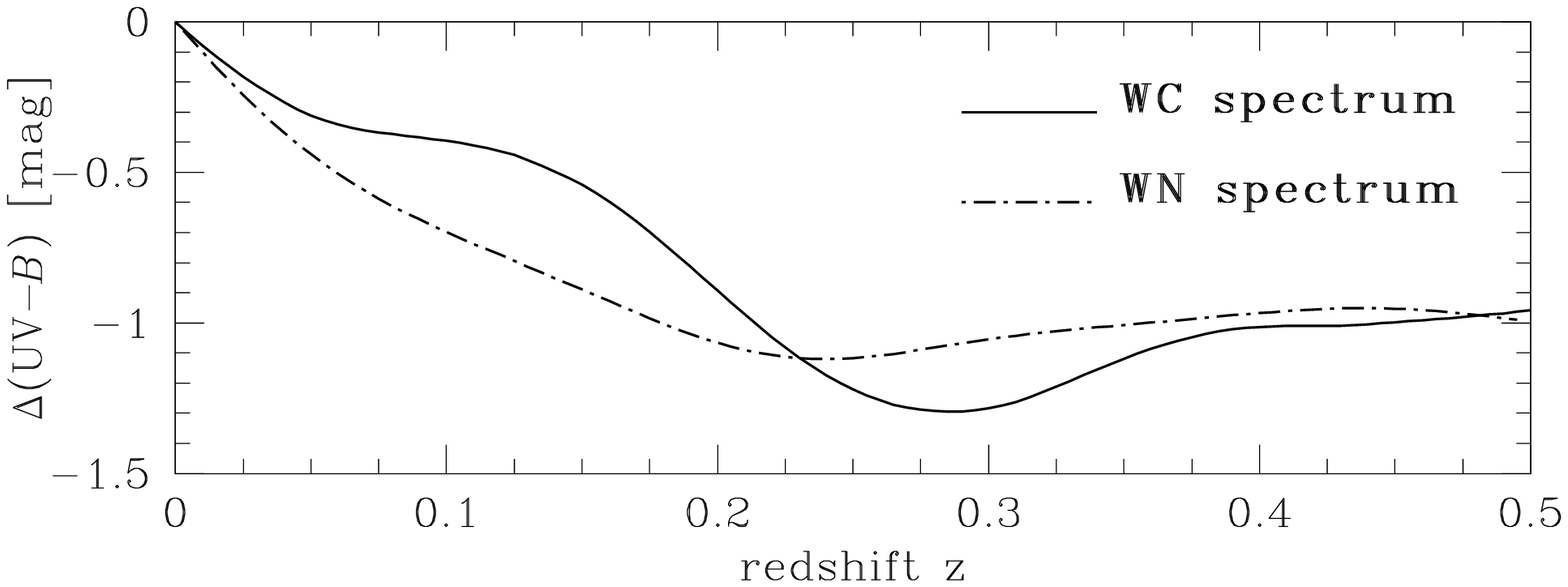}
 
\figcaption[fig4b.ps]{Change in (UV$-B)$ color vs.~redshift, normalized to
$z=0$.  The calculations are based on the HD 165763 WC and the WR152 WN
stellar spectra.}

\noindent
parameterization, $\phi(M) = 0.4 ln10 \phi^*$
$[10^{0.4(M^*-M)}]^{(1+\alpha)}$ $exp(-10^{0.4(M^*-M)})$, where $\phi^*$
is the normalization (galaxies per unit volume), $\alpha$ is the faint end
slope, and $M^*$ is the characteristic absolute magnitude.  To convert the
SSRS2 Sm-Irr $M_B^*=-19.78$ to the UV bandpass, we add the median
(UV$-B)_{\rm o}=-0.91$ color for the UV galaxies.  To simulate WR
galaxies, we add the relative change in (UV$-B)$ over redshift (Figure 4):
 \begin{equation}
M_{\rm UV}^* = M_B^* + ({\rm UV}-B)_{\rm o} + \Delta({\rm UV}-B)[z]
 \end{equation}
	To find the number of galaxies in redshift bin $(z,z+\delta z)$,
we integrate $\phi(M,z)$ to the limiting absolute magnitude $M_{lim}(z)$
over the volume of the redshift interval $z$ to $z+\delta z$. The
selection functions represent the number of galaxies expected per unit
redshift if galaxies are uniformly distributed.

	The curves in Figure 2 are the UV selection functions normalized
to the observed number of galaxies.  The selection function for the S2000
LF ($M^*=-20.59, \alpha=-1.51$) is very similar to the SSRS2 Sm-Irr LF
($M^*=-19.78+($UV$-B)_{\rm o},$ $\alpha=-1.81$), and matches the
(UV$-B)_{\rm o}>-2$ galaxy redshift distribution (Figure 2, upper panel)
quite well.

	In contrast, the UV selection function we calculate for WR
galaxies predicts more WR galaxies at large redshifts.  The redshift
distribution of (UV$-B)_{\rm o}<-2$ galaxies (Figure 2, lower panel)
matches the WR selection function curves much better than the S2000 curve.
The shifts we make in $M^*$ reflect only the change in the flux through
the UV bandpass as a function of redshift.  Thus the UV emission lines of
WR galaxies can explain the extreme blue (UV$-B)_{\rm o}$ colors {\it and}
the redshift distribution of the (UV$-B)_{\rm o}<-2$ galaxies.

	We calculate the number density of Century Survey (which includes
the SA57 region) galaxies to the same effective limiting magnitude and
solid angle as the UV sample, and find that the extreme blue objects are
$\sim$1\% of a ``normal'' ($R$-selected) galaxy population.  It is
difficult to compare the known WR galaxy fraction.  We estimate a lower
limit by assuming the WR catalog of \citet{sch99b} is complete to $B$=14
mag over the northern sky above 15 deg galactic latitude.  The resulting
WR galaxy fraction lower limit is 0.75\%.  Thus the extreme blue galaxy
number density is consistent with the abundance of known WR galaxies.

\section{CONCLUSIONS}

	We show that the extreme blue (UV$-B)_{\rm o}<-2$ galaxies
observed by S2000 may be WR galaxies, starburst galaxies with
strong UV emission from WR stars.  The model explains the observed
(UV$-B)_{\rm o}<-2$ colors and the redshift distribution for the bluest
objects.

	The average (\vr) color of 13 known WR galaxies taken from the
literature, $\overline{(\vr)}=0.27\pm0.34$, is consistent with the range
of (\vr) colors we observe for the extreme blue (UV$-B)_{\rm o}<-2$
galaxies.  We measure a median optical color $(\vr)=0.38\pm0.06$ for 67\%
of the S2000 SA57, UV-selected sample.  

	The UV emission lines can double or triple the flux through the
S2000 UV bandpass at large redshift.  As a result, the (UV$-B)$ color of a
WR galaxy can become up to 1.3 mag bluer at larger redshift.  This
changing (UV$-B)$ as a function of redshift predicts a selection function
weighted toward large redshifts.  Not only do we find an excess of
(UV$-B)_{\rm o}<-2$ galaxies at large redshift, compared to the
(UV$-B)_{\rm o}>-2$ sample, but our model matches the observed redshift
distribution well. We plan to test this model by obtaining high quality
spectra of the suspected WR galaxies and looking for the telltale broad WR
emission features.

\acknowledgements

	We thank R.~Ellis, M.~Sullivan, A.~Barth, C.~Leitherer, and the
referee, H.~Ferguson, for their thoughtful comments and careful reading of
the paper.  We thank A.~Mahdavi for the use of his ``kernel'' code to do
the \citet{pis93} statistical calculation.  This research has made use of
the NASA/IPAC Extragalactic Database (NED) which is operated by the Jet
Propulsion Laboratory, California Institute of Technology, under contract
with NASA.  This research was supported by the Harvard-Smithsonian Center
for Astrophysics.


\begin{thebibliography}{}

\bibitem[Bertin \&Arnouts(1996)]{ber96} Bertin, E., \& Arnouts, S. 1996,
\aap, 117, 393

\bibitem[Calzetti(1997)]{cal97} Calzetti, D. 1997, AIP Conf. Proc., 408,
403

\bibitem[Dufour et al.(1993)]{duf93} Dufour, R. J., Skillman, E. D.,
Garnett, D. R., Shields, G. A., Peimbert, M., Torres-Peimbert, S.,
Terlevish, E., \& Terlevich, R. 1993, RMAA, 27, 115

\bibitem[Fioc \& Rocca-Volmerange(1997)]{fio97} Fioc, M., \&
Rocca-Volmerange, B. 1997, \aap, 362, 950

\bibitem[Folkes et al.(1999)]{fol99} Folkes, S., et al. 1999, \mnras, 308,
459

\bibitem[Frei \& Gunn(1994)]{fre94} Frei, Z., \& Gunn, J. E. 1994, \aj,
108, 1476

\bibitem[Freudling(1993)]{fre93} Freudling, W. 1993, in 5th ESO/ST-ECF
Data Analysis Workshop, ed. P.J. Grosbol \& R.C.E. de Ruijsscher (Garching
bei M\"{u}nchen: ESO), 27

\bibitem[Gallego et al.(1995)]{gal95} Gallego, J., Zamarano, J.,
Arag\'on-Salamanca, A., \& Rego, M. 1995, \apjl, 455, L1-L4

\bibitem[Geller et al.(1997)]{gel97} Geller, M. J., Kurtz, M. J., Wegner,
G., Thorstensen, J. R., Fabricant, D. G., Marzke, R. O., Huchra, J. P.,
Schild, R. E., \& Falco, E. E. 1997, \aj, 114, 2205

\bibitem[Hamann \& Koesterke(1998)]{ham98} Hamann, W.-R., \& Koesterke, L.
1998, \aap, 333, 251

\bibitem[Hillier \& Miller(1999)]{hil99} Hillier, D. \& Miller, D. 1999,
\apj, 519, 354

\bibitem[Kurt et al.(1999)]{kur99} Kurt, C. M., Dufour, R. J., Garnett, D.
R., Skillman, E. D., Mathis, J. S., Peimbert, M., Torres-Peimbert, S., \&
Ruiz, M.-T. 1999, \apj, 518, 246

\bibitem[Landolt(1992)]{lan92} Landolt, A. 1992, \aj, 104, 340

\bibitem[Leitherer et al.(1999)]{lei99} Leitherer C., et al. 1999, \apjs,
123, 3

\bibitem[Lilly et al.(1995)]{lil95} Lilly, S. J., Le F\`evre, O.,
Crampton, D., Hammer, F., \& Tresse, L. 1995, \apj, 455, 108

\bibitem[Kinney et al.(1996)]{kin96} Kinney, A. L., Calzetti, D., Bohlin,
R. C., McQuade, K., Storchi-Bergmann, T., \& Schmitt, H. R. 1996, \apj,
467, 38

\bibitem[Kotilainen et al.(1993)]{kot93} Kotilainen, J. K., Ward, M. J.,
\& Williger, G. M. 1993, \mnras, 263, 655

\bibitem[McQuade et al.(1995)]{mcq95} McQuade, K., Calzetti, D., \&
Kinney, A. L. 1995, \apjs, 97, 331

\bibitem[Madau et al.(1996)]{mad96} Madau, P., Pozzetti, L., Dickinson, M.
E., Giavalisco, M., Steidel, C. C., \& Fruchter, A. 1996, \mnras, 283, 1388

\bibitem[Marzke et al.(1998)]{mar98} Marzke, R. O., da Costa, L. N.,
Pellegrini, P. S., Willmer, C. N., \& Geller, M. J. 1998, \apj, 503, 617

\bibitem[Milliard et al.(1992)]{mil92} Milliard, B., Donas, J., Laget, M.,
Armand, C., \& Vuillemin, A. 1992, \aap, 257, 24

\bibitem[Montgomery, Marschall, \& Janes(1993)]{mon93} Montgomery, K. A.,
Marschall, L. A., \& Janes K. A.  1993, \aj, 106, 181

\bibitem[Pisani(1993)]{pis93} Pisani, A. 1993, \mnras, 265, 706

\bibitem[Schaerer et al.(1999)]{sch99} Schaerer, D., Contini, T.,
\& Kunth, D. 1999, \aap, 341, 399

\bibitem[Schaerer et al.(1999b)]{sch99b} Schaerer, D., Contini, T.,
\& Pindao, M. 1999b, AASS, 136, 35

\bibitem[Schechter(1976)]{sch76} Schechter, P. 1976, \apj, 203, 297

\bibitem[Schlegel, Finkbeiner, \& Davis(1998)]{sch98} Schlegel, D. J.,
Finkbeiner, D. P., \& Davis, M. 1998, \apj, 500, 525

\bibitem[Sullivan et al.(2000)]{S2000} Sullivan, M., Treyer, M. A., Ellis,
R. S., Bridges, T. J., Milliard, B., \& Donas, J. 2000, \mnras, 312, 442

\bibitem[Treyer et al.(1998)]{tre98} Treyer, M. A., Ellis, R. S.,
Milliard, B., Donas, J., \& Bridges, T. J. 1998, \mnras, 300, 303

\bibitem[Willmer et al.(1996)]{wil96} Willmer, C. N. A., Koo D. C.,
Ellman, N., Kurtz, M. J., \& Szalay, A. S. 1996, \apjs, 104, 199

\end{thebibliography}
\end{document}